\documentclass[aps,onecolumn,superscriptaddress]{revtex4}
\usepackage{times}
\usepackage{color}
\usepackage{float}
\usepackage{epsfig}

\begin{document}

\title{Synchronized and mixed outbreaks of coupled recurrent epidemics}

\author{Muhua Zheng}
\affiliation{Department of Physics, East China Normal University,
Shanghai, 200062, P. R. China}
\affiliation{Levich Institute and Physics Department, City College of New York, New York, New York 10031, USA}

\author{Ming Zhao}
\affiliation{College of Physics and Technology, Guangxi Normal University, Guilin 541004, China}

\author{Byungjoon Min}
\affiliation{Levich Institute and Physics Department, City College of New York, New York, New York 10031, USA}

\author{Zonghua Liu}
\email{zhliu@phy.ecnu.edu.cn} \affiliation{Department of Physics,
East China Normal University, Shanghai, 200062, P. R. China}

\begin{abstract}
Epidemic spreading has been studied for a long time and most of them are focused on the growing aspect of
a single epidemic outbreak. Recently, we extended the study to the case of recurrent epidemics (Sci. Rep.
{\bf 5}, 16010 (2015)) but limited only to a single network. We here report from the real data of coupled
regions or cities that the recurrent epidemics in two coupled networks are closely related to each other and
can show either synchronized outbreak pattern where outbreaks occur simultaneously in both networks or mixed
outbreak pattern where outbreaks occur in one network but do not in another one. To reveal the underlying
mechanism, we present a two-layered network model of coupled recurrent epidemics to reproduce the synchronized
and mixed outbreak patterns. We show that the synchronized outbreak pattern is preferred to be triggered in
two coupled networks with the same average degree while the mixed outbreak pattern is likely to show for the
case with different average degrees. Further, we show that the coupling between the two layers tends to
suppress the mixed outbreak pattern but enhance the synchronized outbreak pattern. A theoretical analysis
based on microscopic Markov-chain approach is presented to explain the numerical results. This finding opens a
new window for studying the recurrent epidemics in multi-layered networks.
\end{abstract}

\maketitle

Epidemic spreading in complex networks has been well studied and a lot of great progress have been achieved such
as the infinitesimal threshold \cite{Pastor-Satorras:2001,Boguna:2002,Ferreira:2012,Boguna:2013,Parshani:2010,Castellano:2010},
reaction-diffusion model \cite{Colizza:2007a,Colizza:2007b,Andrea:2008,Liu:2009}, flow driven epidemic \cite{Vazquez:2007,Meloni:2009,Balcan:2009,Ruan:2012,Liu:2012}, and objective spreading etc \cite{Tang:2009,Liu:2010},
see the review Refs. \cite{Pastor:2015,Barrat:2008,Dorogovtsev:2008} for details. Recently, the attention has been
moved to the case of multilayer networks \cite{Boccaletti:2014,Kivela:2014,Feng:2015,Sahneh:2013,Wang:2013,Yagan:2013,
Newman:2005,Marceau:2011,Pastor:2015,Zhao:2014,Buono:2015,Buono:2014}, which represent the interactions
between different real-world networks such as critical infrastructure \cite{little,rosato,buldyrev}, transportation
networks \cite{domenico,parshani}, living organisms \cite{reis,vidal,white}, and social networks \cite{Boccaletti:2014,Min:2016}
etc. These models enable us to determine how the interplay between network structures influences the dynamic processes
taking place on them \cite{Funk:2010,Allard:2009,Dickison:2012,Mendiola:2012,Son:2012,Sanz:2012,Souza:2009,Granell:2013,Cozzo:2013,
Hackett:2016}. For instance, a pathogen spreads on a human contact network abetted by global and regional transportation
networks \cite{Boccaletti:2014}. Due to their ubiquitous applications in complex systems \cite{Halu:2014,Barigozzi:2010,Barigozzi:2011,Cardillo:2013,Kaluza:2010}, the understanding of the properties and dynamic
processes in multilayered networks carries great practical significance.

Two of the most successful models used to describe epidemic spreading are the susceptible-infected-susceptible (SIS)
and susceptible-infected-refractory (SIR) models. Mark \emph{et al} used the SIR model to multilayered networks in 2012
\cite{Dickison:2012}. Very interesting, they found a mixed phase in weakly coupled networks where an
epidemic occurs in one network but does not spread to the coupled network. Saumell-Mendiola \emph{et al} used the SIS
model to multilayered networks also in 2012 \cite{Mendiola:2012}. However, they found that such a mixing phase doesn't
exist in both analytic and simulation results. In their work, they mainly focused on the epidemic threshold and studied
how epidemics spread from one network to another.

All these studies are focused on the growing aspect of a single epidemic outbreak, no matter it is one layer or
multilayered networks. However, in realistic situations, the empirical data shows that epidemic is recurrent, i.e.
with outbreaks from time to time \cite{www:hongkong,Zheng:2015,measles:2016,Scarpino:2016,Panhuis:2013}. Thus, we
recently extended the study to the case of recurrent epidemics \cite{Zheng:2015}, but limited only to a single network.
Considering that the interactions between different networks are ubiquitous, we here recheck several real data of
coupled regions or cities such as the General Out-Patient Clinics (GOPC) network and its coupled General Practitioners
(GP) network of Hong Kong (see Fig. \ref{Fig:data}(a) and (b)), the coupled regional networks of California and Nevada
(see Fig. \ref{Fig:data}(c) and (d)), the coupled regional networks of Arizona and California (see Fig. 1(a) and (b)
in SI), the coupled city networks of Boston and Fall River (see Fig. 1(c) and (d) in SI), and the coupled city networks
of Los Angeles and Sacramento (see Fig. 1(e) and (f) in SI). We interestingly find that their recurrent
epidemics are closely related to each other. Moreover, we find that the coupled time series of recurrent epidemics can
show either synchronized outbreak pattern where outbreaks occur simultaneously in both networks or mixed outbreak pattern
where outbreaks occur only in one network but do not in another one. This finding calls our great interest and
motivates us to study its underlying mechanism. In this sense, we believe that it is very necessary to further extend
the study of recurrent epidemics to the case of multilayered networks.

In this work, we present a two-layered network model of coupled recurrent epidemics to reproduce the synchronized and
mixed outbreak patterns. To guarantee the appearance of recurrent outbreaks, we choose the
susceptible-infected-refractory-susceptible (SIRS) model for each node of network and let the infectious rate be time
dependent, symbolizing the larger annual variation of environment. By this model, we find that the average degrees of
both the intra- and inter-networks play key roles on the emergence of synchronized and mixed outbreak patterns. The
synchronized outbreak pattern tends to be triggered in two coupled networks with the same average degree while the
mixed outbreak pattern is likely to show for the case with different average degrees. Further, we show that the increasing
of coupling strength, i.e. either the inter-layer infection rate or inter-layer average degree, will tend to suppress
the mixed outbreak pattern but enhance the synchronized outbreak pattern. A theoretical analysis
based on microscopic Markov-chain approach is presented to explain the numerical results. This finding may be of
significance to the long-term prediction and control of recurrent epidemics in multi-areas or cities.

\section*{Results}
\begin{figure}
\epsfig{figure=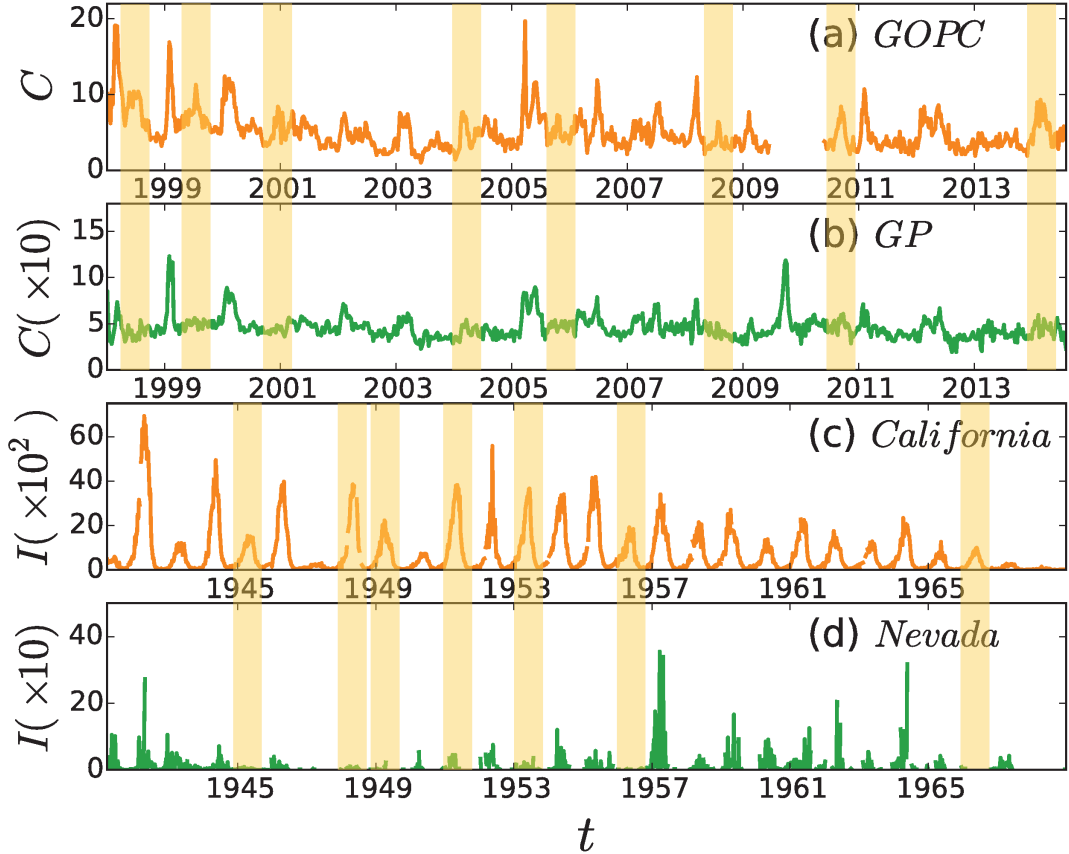,width=0.8\linewidth}
\caption{(color online). {\bf Time series of recurrent epidemics in two coupled regions or cities.}
(a) and (b) represent the weekly consultation rates of influenza-like illness (per 1000 consultations) from $1999$ to
$2013$ in Hong Kong for the General Out-Patient Clinics (GOPC) and the General Practitioners (GP), respectively, where
the data from 2009/6/13 to 2010/5/23 in (a) are not available and the value of $C$ in (b) is from $0$ to $150$. (c) and
(d) represent the time series of reported weekly measles infective cases $I$ in California and Nevada, respectively.
}
\label{Fig:data}
\end{figure}
{\bf The synchronized and mixed outbreak patterns of recurrent epidemics in real data.}
Monitoring epidemic spreading is vital for us to prevent and control infectious diseases. For this purpose, Hong Kong
Department of Health launched a sentinel surveillance system to collect data of infectious diseases, aiming to analyze
and predict the trend of infectious spreading in different regions of Hong Kong. In this system, there are about $64$
General Out-Patient Clinics (GOPC) and $50$ General Practitioners (GP), which form two sentinel surveillance networks
of Hong Kong \cite{www:hongkong}, respectively. By these two networks, we can obtain the weekly consultation rates of
influenza-like illness (per $1000$ consultations). Fig. \ref{Fig:data}(a) and (b) show the collected data from $1999$ to
$2013$ for the GOPC and GP, respectively, where the data from 2009/6/13 to 2010/5/23 in (a) was not collected by Hong
Kong Department of Health and the value of $C$ in (b) is from $0$ to $150$. This weekly consultation rates of
influenza-like illness can well reflect the overall influenza-like illness activity in Hong Kong. From the data in
Fig. \ref{Fig:data}(a) and (b) we easily find that there are intermittent peaks, marking the recurrent outbreaks of
epidemics. By a second check on the data in Fig. \ref{Fig:data}(a) and (b) we interestingly find that some peaks
occur simultaneously in the two networks at the same time, indicating the appearance of synchronized outbreak pattern.
While other peaks appear only in Fig. \ref{Fig:data}(a) but not in Fig. \ref{Fig:data}(b), indicating the
existence of mixed outbreak pattern (see the light yellow shaded areas).

Is this finding of synchronized and mixed outbreak patterns in recurrent epidemics a specific phenomenon only in Hong Kong?
To figure out its generality, we have checked a large number of other recurrent infectious data in different
pathogens and in different states and cities in the United States and found the similar phenomenon. Fig. \ref{Fig:data}(c)
and (d) show the data of weekly measles infective cases $I$ in the states of California and Nevada, respectively, which
were obtained from the USA National Notifiable Diseases Surveillance System as digitized by Project Tycho
\cite{measles:2016,Scarpino:2016,Panhuis:2013}. As California is adjacent to Nevada in west coast of the United States,
their climatic conditions are similar. Thus, they can be also considered as two coupled networks. Three more these kinds
of examples have been shown in Fig. 1 of SI where the coupled networks are based on states-level influenza data and
cities-level measles data, respectively. Therefore, the synchronized and mixed outbreak patterns are general in recurrent
epidemics. We will explain their underlying mechanisms in the next subsection.

{\bf A two-layered network model of coupled recurrent epidemics.}
Two classical models of epidemic spreading are the Susceptible-Infected-Susceptible (SIS) model and
Susceptible-Infected-Refractory (SIR) model \cite{Pastor:2015}. In the SIS model, a susceptible node will be infected by an
infected neighbor with rate $\beta$. In the meantime, each infected node will recover with a probability $\gamma$ at
each time step. After the transient process, the system reaches a stationary state with a constant infected density $\rho_I$.
Similarly, in the SIR model, each node can be in one of the three states: Susceptible, Infected, and Refractory. At each time
step, a susceptible node will be infected by an infected neighbor with rate $\beta$ and an infected node will become
refractory with probability $\gamma$. The infection process will be over when there is no infected $I$. These two models
have been widely used in a variety of situations. However, it was pointed out that both the SIS and SIR models are failed to
explain the recurrence of epidemics in real data \cite{Zheng:2015,Stone:2007}.
\begin{figure}
\epsfig{figure=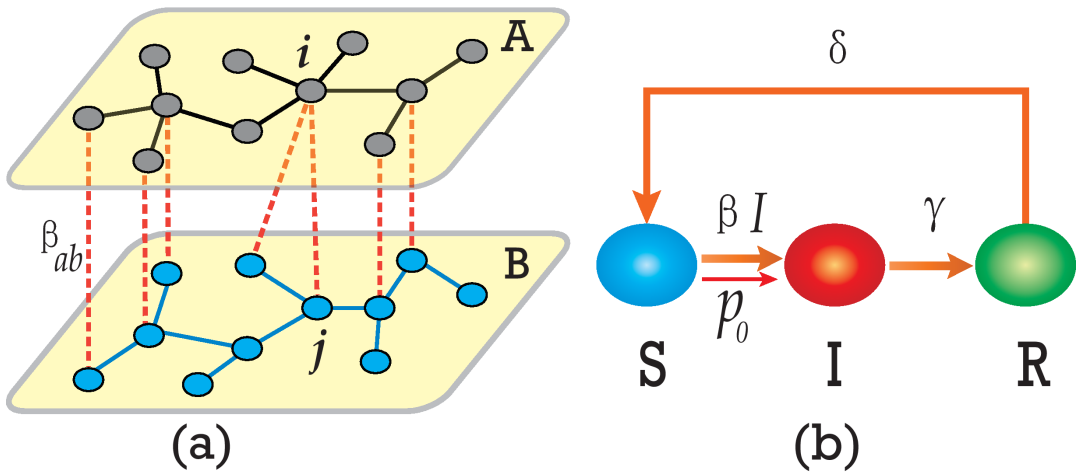,width=0.8\linewidth}
\caption{(color online). {\bf Schematic figure of the epidemic model to reproduce the synchronized and mixed outbreak patterns}.
(a) Schematic figure of the two-layered network, where the ``black", ``blue" and ``red" lines represent the links of the
networks $\mathcal{A}$, $\mathcal{B}$ and the inter-network $\mathcal{AB}$, respectively. $\beta_{ab}$ denotes the infectious
rate through one interconnection between $\mathcal{A}$ and $\mathcal{B}$. (b) Schematic figure of the extended SIRS model for
each node in $\mathcal{A}$ and $\mathcal{B}$, where the symbols $S, I$ and $R$ represent the susceptible, infectious, and
refractory states, respectively, and the parameters $\beta, \gamma$ and $\delta$ represent the infectious, refractory and
recovery rates, respectively. $p_0$ denotes the probability for a susceptible node to be naturally infected by environment or
other factors. }
\label{Fig:model}
\end{figure}

To reproduce the synchronized and mixed outbreak patterns, we here present a two-layered network model of coupled recurrent
epidemics, shown in Fig. \ref{Fig:model} where (a) represents its schematic figure of network topology and (b) denotes the
epidemic model at each node. In Fig. \ref{Fig:model}(a), two networks $\mathcal{A}$ and $\mathcal{B}$ are coupled through
some interconnections between them, which form the inter-network $\mathcal{AB}$. For the sake of simplicity, we let the
two networks $\mathcal{A}$ and $\mathcal{B}$ have the same size $N_a=N_b$. We let $\langle k_a\rangle$, $\langle k_b\rangle$,
and $\langle k_{ab}\rangle$ represent the average degrees of the networks $\mathcal{A}$, $\mathcal{B}$ and$\mathcal{AB}$,
respectively, see {\sl Methods} for details. In Fig. \ref{Fig:model}(b), the epidemic model is adopted from our previous
work \cite{Zheng:2015} by two steps. In step one, we extend the SIR model to a Susceptible-Infected-Refractory-Susceptible
(SIRS) model where a refractory node will become susceptible again with probability $\delta$. In step two, we let each
susceptible node have a small probability $p_0$ to be infected, which represents the natural infection from
environment. Moreover, we let the infectious rate $\beta(t)$ be time dependent, representing its annual and seasonal
variations etc. To distinguish the function of the interconnections from that of those links in $\mathcal{A}$ and
$\mathcal{B}$, we let $\beta_{ab}$ be the inter-layer infectious rate. In this way, the interaction between $\mathcal{A}$
and $\mathcal{B}$ can be described by the tunable parameter $\beta_{ab}$.

\begin{figure}
\epsfig{figure=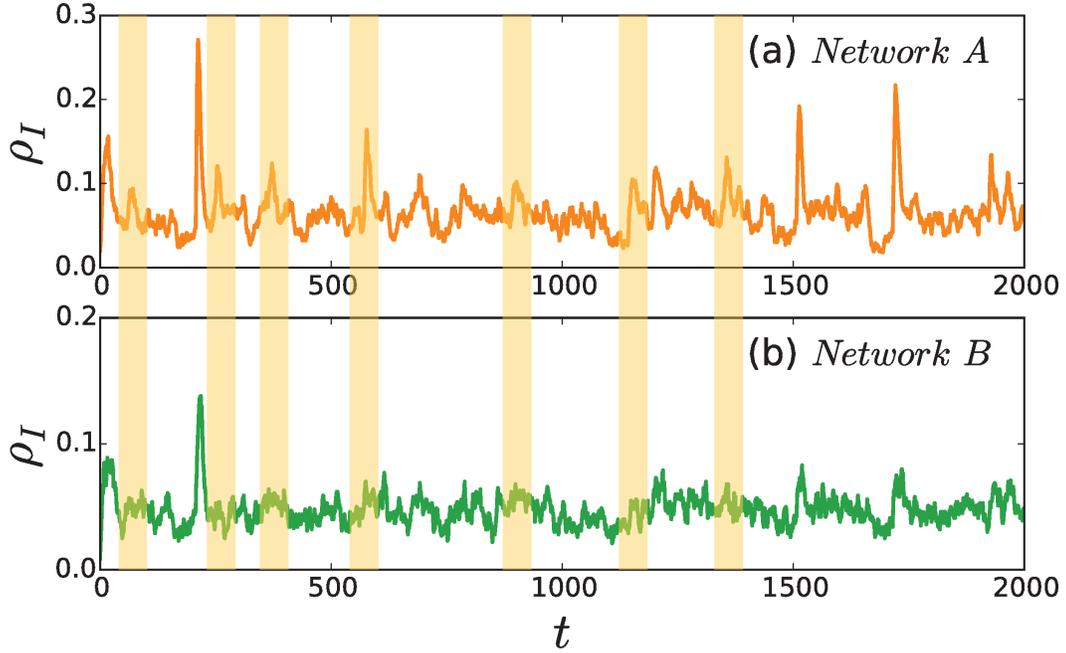,width=0.8\linewidth}
\caption{(color online). {\bf Reproduced time series of recurrent epidemics in the two-layered network model.}
(a) and (b) represent the evolutions of the infected density $\rho_I$ in the two-layered networks $\mathcal{A}$ and
$\mathcal{B}$, respectively, where the parameters are taken as $\langle k_a\rangle=6.5$, $\langle k_b\rangle=1.5$,
$\langle k_{ab}\rangle=1.0$, $\beta(t) \sim \mathcal{N}(0.1,0.1^2)$, $\beta_{ab}=0.09$, $\gamma=0.2$, $\delta=0.02$,
$p_0=0.01$, and $N_a=N_b=1000$.
}
\label{Fig:model_timeseeries}
\end{figure}

In numerical simulations, we let both the networks $\mathcal{A}$ and $\mathcal{B}$ be the Erd\H{o}s-R\'{e}nyi (ER) random
networks \cite{BA:2002}. To guarantee a recurrent epidemics in each of $\mathcal{A}$ and $\mathcal{B}$, we follow
Ref. \cite{Zheng:2015} to let $\beta(t)$ be the truncated Gaussian distribution $\mathcal{N}(0.1,0.1^2)$ and choose
$p_0=0.01$. Fig. \ref{Fig:model_timeseeries}(a) and (b) show the evolutions of the infected density $\rho_I$ in $\mathcal{A}$
and $\mathcal{B}$, respectively, where the parameters are taken as $\langle k_a\rangle=6.5$, $\langle k_b\rangle=1.5$,
$\langle k_{ab}\rangle=1.0$, and $\beta_{ab}=0.09$. It is easy to observe that some peaks of $\rho_I$ appear simultaneously
in $\mathcal{A}$ and $\mathcal{B}$, indicating the synchronized outbreak pattern. We also notice that some peaks of $\rho_I$
in $\mathcal{A}$ do not have corresponding peaks in $\mathcal{B}$ (see the light yellow shadowed areas in
Fig. \ref{Fig:model_timeseeries}(a) and (b)), indicating the mixed outbreak pattern. Moreover, we do not find the contrary
case where there are peaks of $\rho_I$ in $\mathcal{B}$ but no corresponding peaks in $\mathcal{A}$, which is also consistent
with the empirical observations in Fig. \ref{Fig:data}(a)-(d).

{\bf Mechanism of the synchronized and mixed outbreak patterns.}
To understand the phenomenon of synchronized and mixed outbreak patterns better, we here study their underlying mechanisms.
A key quantity for the phenomenon is the outbreak of epidemic, i.e. the peaks in the time series of Fig. \ref{Fig:data}.
Notice that a peak is usually much higher than its background oscillations. To pick out a peak, we need to define its
background/baseline first. As the distributions of both the real data in Fig. \ref{Fig:data} and numerical simulations
in Fig. \ref{Fig:model_timeseeries} are approximately satisfied the normal distribution (see Fig. 2 SI), we define the
baseline as $\mu+ 3\sigma$ with $\mu$ and $\sigma$ being the mean and standard deviation, respectively, which contains
about $99.7\%$ data in the normal distribution \cite{Patel:1996}. Then, we can count the number of outbreaks in a
measured time $t$. Let $n$ be the average number of outbreaks in realizations of the same evolution $t$. Larger $n$
implies more frequent outbreaks. Let $\Delta n=\mid n(\mathcal{A})- n(\mathcal{B})\mid$ be the difference of outbreak
numbers between the networks $\mathcal{A}$ and $\mathcal{B}$. Larger $\Delta n$ implies more frequent emergence of the
mixed outbreak pattern. In particular, the mixed outbreak pattern will disappear when $\Delta n=0$.
\begin{figure}[ht]
\centering
\includegraphics[width=0.80\linewidth]{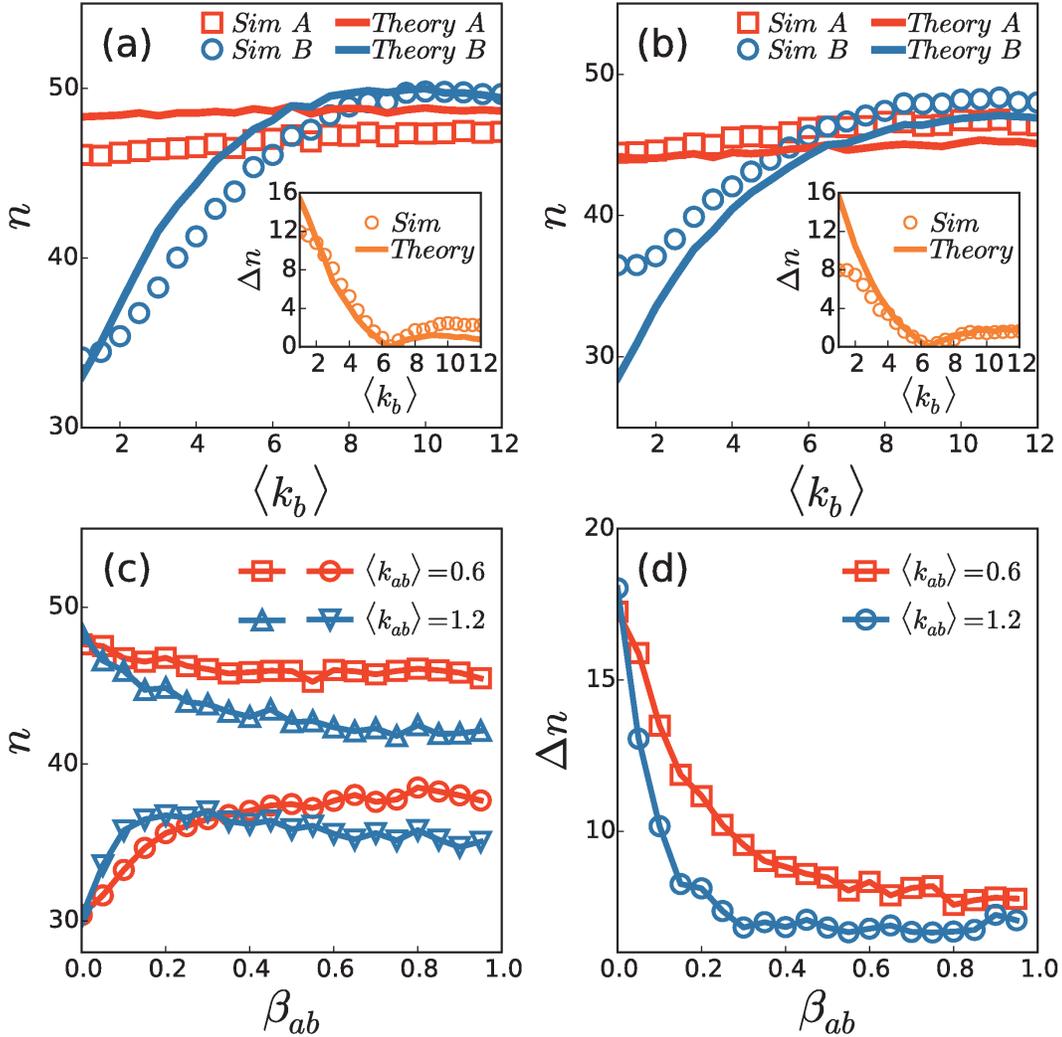}
\caption{(color online).
(a) and (b): Dependence of $n$ on the average degree $\langle k_b \rangle$ of network $\mathcal{B}$ for $\beta_{ab}=0.09$
and $0.3$, respectively, where the average degree of network $\mathcal{A}$ is fixed as $\langle k_a\rangle=6.5$,
$\langle k_{ab}\rangle=1.0$, and the insets show the dependence of
$\Delta n=\mid n(\mathcal{A})- n(\mathcal{B})\mid$ on $\langle k_b \rangle$. The solid lines represent the theoretical
results from Eqs. (\ref{eq:theoryA}) and (\ref{eq:theoryB}).
(c): Dependence of $n$ on $\beta_{ab}$ with $\langle k_a\rangle=6.5$ and $\langle k_b\rangle=1.5$ where the ``squares" and
``circles" represent the case of $\langle k_{ab}\rangle=0.6$ for the networks $\mathcal{A}$ and $\mathcal{B}$, respectively,
and the ``up triangles" and ``down triangles" represent the case of $\langle k_{ab}\rangle=1.2$ for the networks $\mathcal{A}$
and $\mathcal{B}$, respectively.
(d): Dependence of $\Delta n$ on $\beta_{ab}$ with $\langle k_a\rangle=6.5$ and $\langle k_b\rangle=1.5$ where the
``squares" and ``circles" represent the cases of $\langle k_{ab}\rangle=0.6$ and $1.2$, respectively.
The other parameters are set as $\beta(t) \sim \mathcal{N}(0.1,0.1^2)$, $\gamma=0.2$, $\delta=0.02$, $p_0=0.01$, and
$N_a=N_b=1000$. All the results are averaged over $1000$ independent realizations with the simulation
steps $t=20000$.
}
\label{Fig:outbreakB}
\end{figure}

We are interested in how the average degrees and coupling influence the numbers $n$ and $\Delta n$. Figure \ref{Fig:outbreakB}(a)
and (b) show the dependence of $n$ on the average degree $\langle k_b \rangle$ of network $\mathcal{B}$ for $\beta_{ab}=0.09$
and $0.3$, respectively, where the average degree of network $\mathcal{A}$ is fixed as $\langle k_a\rangle=6.5$. It is easy
to see from Fig. \ref{Fig:outbreakB}(a) and (b) that the number $n(\mathcal{B})$ of network $\mathcal{B}$
will gradually increase with the increase of $\langle k_b \rangle$, while the number $n(\mathcal{A})$ of network $\mathcal{A}$
keeps approximately constant, indicating that a larger $\langle k_b \rangle$ is in favor of the recurrent outbreaks.
Specifically, $n(\mathcal{B})$ will reach $n(\mathcal{A})$ when $\langle k_b \rangle$ is increased to the value of
$\langle k_b \rangle=\langle k_a\rangle=6.5$, see the insets in Fig. \ref{Fig:outbreakB}(a) and (b) for the minimum of
$\Delta n$. For details, Fig. 3 in SI shows the evolution of infected densities $\rho_I$ for the cases of
$\langle k_a\rangle=\langle k_b\rangle$, confirming the result of $\Delta n=0$ in Fig. \ref{Fig:outbreakB}(a) and (b). On
the other hand, comparing the two insets in Fig. \ref{Fig:outbreakB} (a) and (b), we find that $\Delta n$ in
Fig. \ref{Fig:outbreakB}(a) is greater than the corresponding one in Fig. \ref{Fig:outbreakB}(b), indicating that a larger
$\beta_{ab}$ is in favor of suppressing the mixed outbreak pattern. These results can be also theoretically explained by the
microscopic Markov-chain approach, see {\sl Methods} for details. The solid lines in Fig. \ref{Fig:outbreakB} (a) and (b)
represent the theoretical results from Eqs. (\ref{eq:theoryA}) and (\ref{eq:theoryB}). It is easy to see that the theoretical
results are consistent with the numerical simulations very well.

Fig. \ref{Fig:outbreakB}(c) and (d) show the influences of the coupling parameters such as $\beta_{ab}$ and $k_{ab}$ for the
outbreak number $n$ and the difference of outbreak number $\Delta n$, respectively, where $\langle k_a\rangle=6.5$ and
$\langle k_b\rangle=1.5$. From Fig. \ref{Fig:outbreakB}(c) we see that for the case of $k_{ab}=0.6$, $n(\mathcal{A})$ is an
approximate constant and $n(\mathcal{B})$ gradually increase with $\beta_{ab}$. While for the case of $k_{ab}=1.2$, both
$n(\mathcal{A})$ and $n(\mathcal{B})$ change with $\beta_{ab}$, indicating that both $k_{ab}$ and $\beta_{ab}$ take important
roles in the synchronized and mixed outbreak patterns. From Fig. \ref{Fig:outbreakB}(d) we see that the case of $k_{ab}=1.2$
decreases faster than the case of $k_{ab}=0.6$, indicating that both the larger $k_{ab}$ and larger $\beta_{ab}$ are in favor
of the synchronized outbreak pattern. That is, the stronger coupling will suppress the mixed outbreak pattern but enhance the
synchronized outbreak pattern. On the contrary, the weaker coupling is in favor of the mixed epidemic outbreak pattern
but suppress the synchronized outbreak pattern. For details, Fig. 4 in SI shows the evolution of infected densities $\rho_I$
for different $\beta_{ab}$, confirming the above results.

\begin{figure*}[ht]
\begin{center}
\centerline{\includegraphics[width=0.8\textwidth]{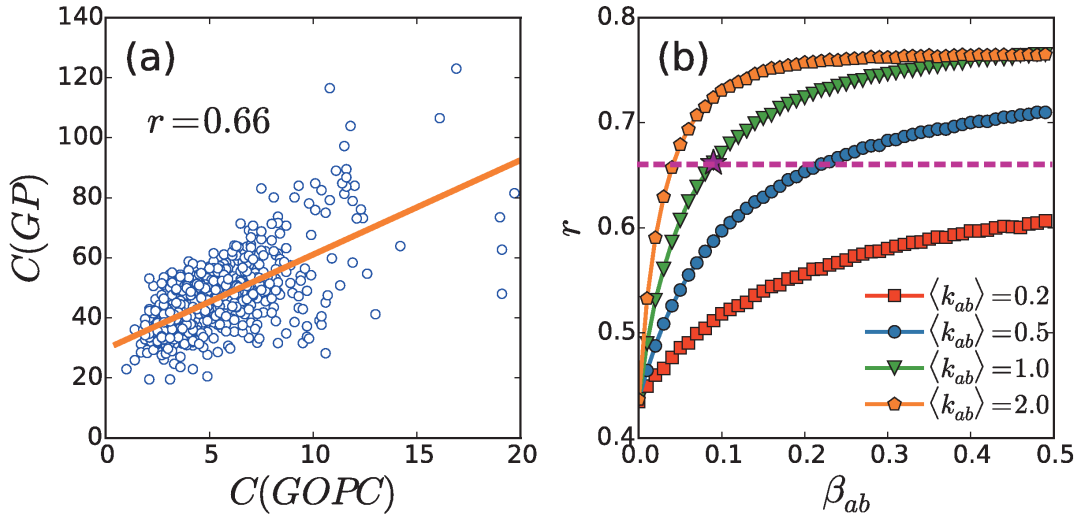}}
\caption{(color online.) {\bf Correlation between the two coupled layers}. (a) Correlation between the two time series of GOPC
and GP. (b) Dependence of the correlation coefficient $r$ on $\beta_{ab}$ for $\langle k_{ab}\rangle=0.2, 0.5, 1.0$ and $2.0$,
respectively, where the purple ``star'' and its related dashed line represent the point of $r=0.66$. Other parameters are the
same as in Fig. \ref{Fig:model_timeseeries}(a) and (b).
}
\label{Fig:correlation}
\end{center}
\end{figure*}

{\bf Coupling induced correlation between the epidemics of the two networks.}
The coupling between the two layers is represented by the pair of variables $(\beta_{ab}, \langle k_{ab}\rangle)$.
Qualitatively, larger $\beta_{ab}$ and $\langle k_{ab}\rangle$ represent stronger coupling. To quantitatively represent the
effects of $\beta_{ab}$ and $\langle k_{ab}\rangle$, we here measure the cross-correlation, $r$, defined in
Eq. (\ref{eq:Correlation coefficient}), which can show some information beyond the synchronized and mixed outbreak patterns.
By Eq. (\ref{eq:Correlation coefficient}) we first calculate the coefficient $r$ between the two time series of GOPC and GP
and find $r=0.66$, indicating that these two data are highly correlated. Fig. \ref{Fig:correlation}(a) shows the
correspondence between the two time series of GOPC and GP. Then, we check the influence of coupling on the coefficient $r$.
Fig. \ref{Fig:correlation}(b) shows the dependence of $r$ on $\beta_{ab}$ for $\langle k_{ab}\rangle=0.2, 0.5, 1.0$ and $2.0$,
respectively. We see that $r$ increases with $\beta_{ab}$ for a fixed $\langle k_{ab}\rangle$ and also increases with
$\langle k_{ab}\rangle$ for a fixed $\beta_{ab}$, indicating the enhanced correlation by the coupling strength. Very
interesting, we find that for the case of $\langle k_{ab}\rangle=1.0$ in Fig. \ref{Fig:correlation}(b), the point of $r=0.66$
corresponds to $\langle \beta_{ab}\rangle=0.09$ (see the purple ``star''), implying that the coupling in
Fig. \ref{Fig:correlation}(a) is equivalent to the case of $\langle k_{ab}\rangle=1.0$ and $\langle \beta_{ab}\rangle=0.09$
in Fig. \ref{Fig:correlation}(b). In this sense, we may draw a horizontal line passing the purple ``star'' in
Fig. \ref{Fig:correlation}(b) (see the dashed line) and its crossing points with all the curves there will also represent the
equivalent coupling in Fig. \ref{Fig:correlation}(a). However, we notice that the horizontal line has no crossing point with
the curve of $\langle k_{ab}\rangle=0.2$, indicating that there is a threshold of $\langle k_{ab}\rangle$ for the appearance
of $r=0.66$ in Fig. \ref{Fig:correlation}(a).

It is maybe helpful to understand the influence of $(\beta_{ab}, \langle k_{ab}\rangle)$ on epidemics from the aspect of
purely coupling in network structure. We know that a critical problem in coupled networks is when they behave as separate
networks vs when they behave as a solid single network \cite{Radicchi:2013,Sahneh:2015}. In our case here, the synchrony
of epidemic in the two layers corresponds the case of strong coupling while the mixed pattern represents the case of weak
coupling. Therefore, the synchrony and mixed patterns also reflect the influence of network structure on its dynamics. On
the other hand, we notice from Fig. \ref{Fig:correlation}(b) that there is a finite value of $r\approx 0.43$ when
$\beta_{ab}=0$, implying that the same infection probability $\beta(t)$ in the two networks do have a contribution to the
correlation $r$. However, the further increase of $r$ in Fig. \ref{Fig:correlation}(b) reflects the influence from the
coupling parameters of $(\beta_{ab}, \langle k_{ab}\rangle)$. From Fig. \ref{Fig:correlation}(b) we see that for a
fixed $\beta_{ab}$, $r$ increases with $\langle k_{ab}\rangle$, indicating the influence of $\langle k_{ab}\rangle$. While
for a fixed $\langle k_{ab}\rangle$, $r$ increases with $\beta_{ab}$, indicating the influence of $\beta_{ab}$. Therefore,
the further increase of correlation from $r\approx 0.43$ do come from the coupling between the two time series.

\section*{Discussion}

The dependence of the synchronized and mixed outbreak patterns on the average degrees of networks $\mathcal{A}$ and $\mathcal{B}$
may be also understood from the aspect of their epidemic thresholds. By the theoretical analysis in {\it Methods} we obtain the
epidemic thresholds as $\beta_c^A=\frac{\gamma}{\langle k_a\rangle}$ and $\beta_c^B=\frac{\gamma}{\langle k_b\rangle}$ in
Eq. (\ref{eq:critical}), when the two networks are weakly coupled. For the case of $\langle k_a\rangle > \langle k_b\rangle$,
we have $\beta_c^A < \beta_c^B$. When $\beta$ satisfies $\beta< \beta_c^A < \beta_c^B$, the epidemics cannot survive in each of
networks $\mathcal{A}$ and $\mathcal{B}$. Thus, the infected fraction will be approximately zero, i.e. no epidemic outbreak
in both $\mathcal{A}$ and $\mathcal{B}$. It should be noticed that this result just holds for the case of weakly coupling. When
coupling is strong, it is possible for epidemic to occur in the coupled network even when $\beta$ is below the epidemic
threshold of each layer \cite{Sahneh:2013}. When $\beta$ satisfies $\beta> \beta_c^B> \beta_c^A$, the epidemics will survive in
both networks $\mathcal{A}$ and $\mathcal{B}$, indicating an outbreak will definitely occur in both of them. These two cases
are trivial. However, when $\beta$ is in between $\beta_c^A$ and $\beta_c^B$, some interesting results may be induced by coupling.
When coupling is weak, it is possible for epidemic to outbreak only in network $\mathcal{A}$ but not in network $\mathcal{B}$. When coupling is
increased slightly, it will be also possible for epidemic to outbreak sometimes in network $\mathcal{B}$, i.e. resulting a mixed outbreak
pattern. Once coupling is further increased to large enough, a synchronized outbreak pattern will be resulted.

So far, the reported results are from the ER random networks $\mathcal{A}$ and $\mathcal{B}$. We are wondering whether it is
possible to still observe the phenomenon of the synchronized and mixed outbreak patterns in other network topologies. For this
purpose, we here take the scale-free network \cite{BA:2002} as an example. Very interesting, by repeating the above simulation
process in scale-free networks we have found the similar phenomenon as in ER random networks, see Figs. 5-7 in SI for details.
Therefore, the synchronized and mixed outbreak patterns are a general phenomena in multi-layered epidemic networks.

In sum, the epidemic spreading has been well studied in the past decades, mainly focused on both the single and multi-layered
networks. However, only a few works focused on the aspect of recurrent epidemics, including both the models in homogeneous
population \cite{Stone:2007} and our recent model in a single network \cite{Zheng:2015}. We here report from the real data that
the epidemics from different networks are in fact not isolated but correlated, implying that they should be considered as a
multi-layered network. Motivated by this discovery, we present a two-layered network model to reproduce the correlated recurrent
epidemics in coupled networks. More importantly, we find that this model can reproduce both the synchronized and mixed outbreak
patterns in real data. The two-layered network favors to show the synchronized pattern when the average degrees of the two
coupled networks have a large difference and shows the mixed pattern when their average degrees are very close. Besides the degree
difference between the two networks, the coupling strength between the two layers has also significant influence to the
synchronized and mixed outbreak patterns. We show that both the larger $\beta_{ab}$ and larger $\langle k_{ab}\rangle$ are in
favor of the synchronized pattern but suppress the mixed pattern. This finding thus shows a new way to understand the epidemics in
realistic multi-layered networks. Its further studies and potential applications in controlling the recurrent epidemics may be
an interesting topic in near future.

\section*{Methods}

\subsection*{A two-layered network model of recurrent epidemic spreading}
We consider a two-layered network model with coupling between its two layers, i.e. the networks $\mathcal{A}$ and $\mathcal{B}$
in Fig. \ref{Fig:model}(a). We let the two networks have the same size $N_a=N_b=N$ and their degree distributions $P_A(k)$ and
$P_B(k)$ be different. We may imagine the network $\mathcal{A}$ as a human contact network for one geographic region and the
network $\mathcal{B}$ for a separated region. Each node in the two-layered network has two kinds of links, i.e. intra-links 
within $\mathcal{A}$ or $\mathcal{B}$  and the interconnection between $\mathcal{A}$ and $\mathcal{B}$. The former consists of the degree
distributions $P_A(k)$ and $P_B(k)$ while the latter the interconnection network. We let $\langle k_a\rangle$,
$\langle k_b\rangle$, and $\langle k_{ab}\rangle$ represent the average degrees of networks $\mathcal{A}$, $\mathcal{B}$ and
interconnection network $\mathcal{AB}$, respectively. In details, we firstly generate two separated networks $\mathcal{A}$ and
$\mathcal{B}$ with the same size $N$ and different degree distributions $P_A(k)$ and $P_B(k)$, respectively. Then, we add links
between $\mathcal{A}$ and $\mathcal{B}$. That is, we randomly choose two nodes from $\mathcal{A}$ and $\mathcal{B}$ and then
connect them if they are not connected yet. Repeat this process until the steps we planned. In this way, we obtain an
uncorrelated two-layered network.

To discuss epidemic spreading in the two-layered network, we use the extended SIRS model, see Fig. \ref{Fig:model}(b)
for its schematic illustration. In this model, a susceptible node has three ways to be infected. The first one is the natural
infection from environment or unknown reasons, represented by a small probability $p_0$. The second one is the infection from
contacting with infected individuals in the network $\mathcal{A}$ (or $\mathcal{B}$), represented by $\beta(t)$. And the third
one is the infection from another network, represented by $\beta_{ab}$ (see Fig. \ref{Fig:model}(a)). Thus, a susceptible node
will become infected with a probability $1-(1-p_0)(1-\beta(t))^{k^{inf}}(1-\beta_{ab})^{k_{ab}^{inf}}$ where $k^{inf}$ is the
infected neighbors in the same network and $k_{ab}^{inf}$ is the infected neighbors in another network. At the same time, an
infected node will become refractory by a probability $\gamma$ and a refractory node will become susceptible again by a
probability $\delta$.

In numerical simulations, the dependence of $\beta(t)$ on time is implemented as follows \cite{Zheng:2015}: we divide the total
time $t$ into multiple segments with length $T$ and let $T=52$, corresponding to the $52$ weeks in one year. We let
$\beta(t)$ be a constant in each segment, which is randomly chosen from the truncated Gaussian distribution $\mathcal{N}(0.1,0.1^2)$.
Once a $\beta(t)<0$ or $\beta(t)>1$ is chosen, we discard it and then choose a new one. At the same time, we fix $\gamma=0.2$
and $\delta=0.02$ and set $\beta_{ab}$ as the tunable parameter.

\subsection*{A theoretical analysis based on microscopic Markov-chain approach}
\begin{figure*}[ht]
\begin{center}
\centerline{\includegraphics[width=0.9\textwidth]{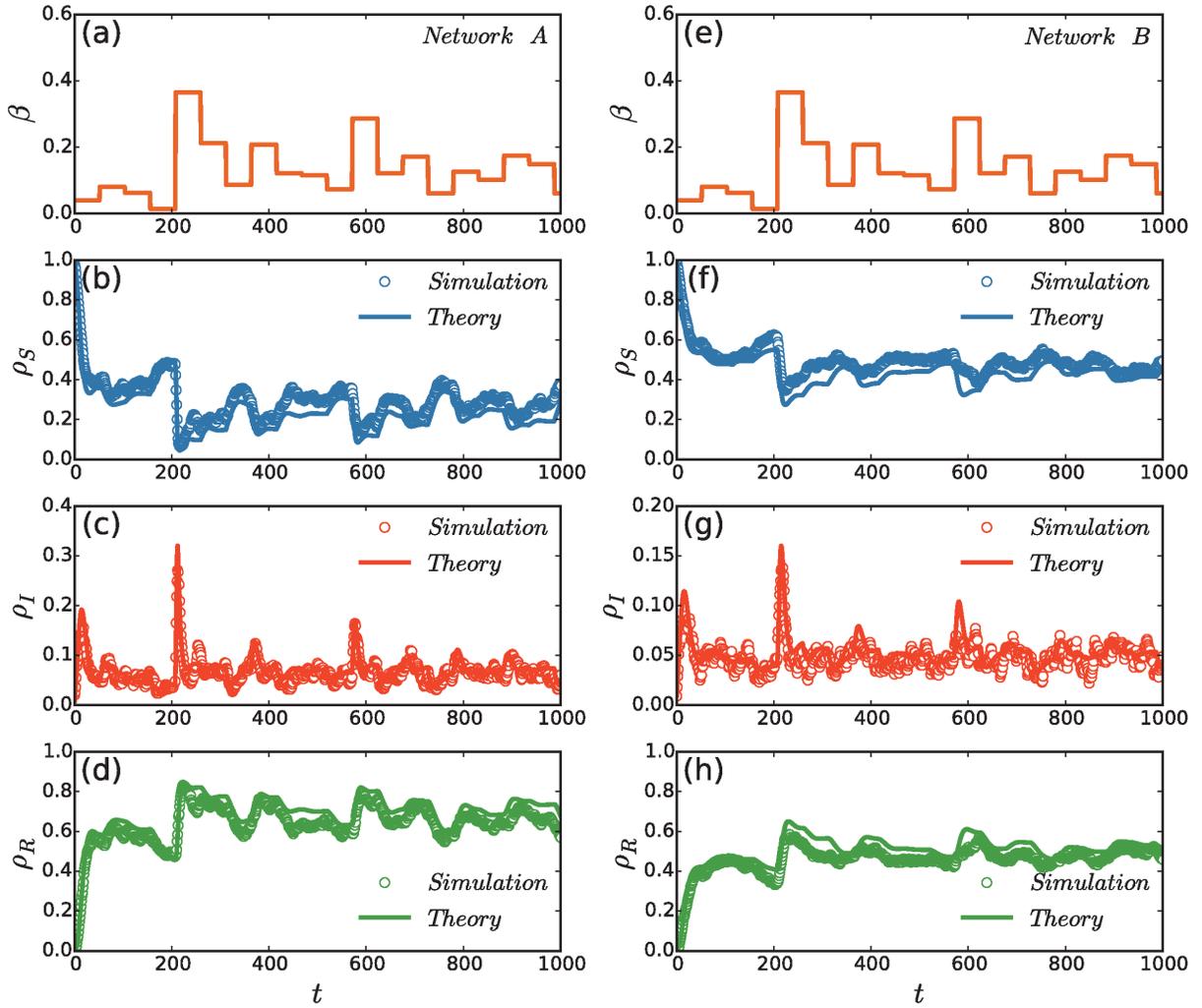}}
\caption{(color online.) {\bf Comparison between the theoretical solutions and numerical simulations.}
The left and right panels are for the networks $\mathcal{A}$ and $\mathcal{B}$, respectively. All the parameters are the same
as in Fig. \ref{Fig:model_timeseeries}(a) and (b).
(a) and (e) $\beta(t)$ versus $t$; (b) and (f) $\rho_S$ versus $t$; (c) and (g) $\rho_I$ versus $t$; (d) and (h) $\rho_R$ versus
$t$. In (b)-(d) and (f)-(h), the solid curves represent the theoretical solutions while the ``circles" represent the numerical
simulations.
}
\label{Fig:theory}
\end{center}
\end{figure*}

Let $P^S_{i,A}(t)$, $P^I_{i,A}(t)$, $P^R_{i,A}(t)$ be the probabilities for node $i$ in network $\mathcal{A}$ to be in one of the
three states of $S, I$ and $R$ at time $t$, respectively. Similarly, we have $P^S_{i,B}(t)$, $P^I_{i,B}(t)$ and $P^R_{i,B}(t)$ in
network $\mathcal{B}$. They satisfy the conservation law
\begin{eqnarray}\label{eq:normalization}
P^S_{i,A}(t)+P^I_{i,A}(t)+P^R_{i,A}(t)=1,\quad
P^S_{i,B}(t)+P^I_{i,B}(t)+P^R_{i,B}(t)=1.
\end{eqnarray}
According to the Markov-chain approach \cite{Gomez:2010,Gomez:2011,Granell:2013,Zheng:2015,Valdano:2015}, we introduce
\begin{eqnarray*}
\rho^A_S(t)=\frac{1}{N}\sum\limits_{i=1}^NP^S_{i,A}(t) ,\quad
\rho^A_I(t)=\frac{1}{N}\sum\limits_{i=1}^NP^I_{i,A}(t) ,\quad
\rho^A_R(t)=\frac{1}{N}\sum\limits_{i=1}^NP^R_{i,A}(t) ,
\end{eqnarray*}
\begin{eqnarray*}
\rho^B_S(t)=\frac{1}{N}\sum\limits_{i=1}^NP^S_{i,B}(t) ,\quad
\rho^B_I(t)=\frac{1}{N}\sum\limits_{i=1}^NP^I_{i,B}(t) ,\quad
\rho^B_R(t)=\frac{1}{N}\sum\limits_{i=1}^NP^R_{i,B}(t) ,
\end{eqnarray*}
where $\rho^A_S(t)$, $\rho^A_I(t)$ and $\rho^A_R(t)$ represent the densities of susceptible, infected, and refractory
individuals at time $t$ in network $\mathcal{A}$, respectively. Similarly, we have $\rho^B_S(t)$, $ \rho^B_I(t)$ and
$\rho^B_R(t)$ in network $\mathcal{B}$.

Let $q^{S,I}_{i,A}(t)$, $q^{I,R}_{i,A}(t)$ and $q^{R,S}_{i,A}(t)$ be the transition probabilities from the
state $S$ to $I$, $I$ to $R$, and $R$ to $S$ in network $\mathcal{A}$, respectively. By the Markov chain approach
\cite{Gomez:2010,Valdano:2015} we have
\begin{eqnarray}\label{eq:qsiA}
q^{S,I}_{i,A}(t)&=&1-(1-p_0)\prod_{l\in\Lambda_{i,A}}(1-\beta(t)P^I_{l,A}(t))\prod_{v\in \Lambda_{i,B}}(1-\beta_{ab}P^I_{v,B}(t)),\nonumber \\
q^{I,R}_{i,A}(t)&=&\gamma, \\
q^{R,S}_{i,A}(t)&=&\delta, \nonumber
\end{eqnarray}
where $\Lambda_{i,A}$ represents the neighboring set of node $i$ in network $\mathcal{A}$. The term $(1-p_0)$ in
Eq. (\ref{eq:qsiA}) represents the probability that node $i$ is not infected by the environment. The term
$\prod_{l\in\Lambda_{i,A}}(1-\beta(t)P^I_{l,A}(t))$ is the probability that node $i$ is not infected by the infected neighbors
in network $\mathcal{A}$. While the term $\prod_{v\in \Lambda_{i,B}}(1-\beta_{ab}P^I_{v,B}(t))$ is the probability that node
$i$ is not infected by the infected neighbors in another network. Thus,
$(1-p_0)\prod_{l\in\Lambda_{i,A}}(1-\beta(t)P^I_{l,A}(t))\prod_{v\in \Lambda_{i,B}}(1-\beta_{ab}P^I_{v,B}(t))$ is the
probability for node $i$ to be in the susceptible state. Similarly, for the node in network $\mathcal{B}$, we have
\begin{eqnarray}\label{eq:qsiB}
q^{S,I}_{i,B}(t)&=&1-(1-p_0)\prod_{l\in\Lambda_{i,B}}(1-\beta(t)P^I_{l,B}(t))\prod_{v\in \Lambda_{i,A}}(1-\beta_{ab}P^I_{v,A}(t)),\nonumber \\
q^{I,R}_{i,B}(t)&=&\gamma, \\
q^{R,S}_{i,B}(t)&=&\delta. \nonumber
\end{eqnarray}

Based on these analysis, we formulate the following difference equations
\begin{eqnarray}\label{eq:MC1}
P^S_{i,A}(t+1)&=&P^S_{i,A}(t)(1-q^{S,I}_{i,A}(t))+P^R_{i,A}(t)q^{R,S}_{i,A}(t),\nonumber \\
P^I_{i,A}(t+1)&=&P^I_{i,A}(t)(1-q^{I,R}_{i,A}(t))+P^S_{i,A}(t)q^{S,I}_{i,A}(t), \\
P^R_{i,A}(t+1)&=&P^R_{i,A}(t)(1-q^{R,S}_{i,A}(t))+P^I_{i,A}(t)q^{I,R}_{i,A}(t). \nonumber
\end{eqnarray}
\begin{eqnarray}\label{eq:MC2}
P^S_{i,B}(t+1)&=&P^S_{i,B}(t)(1-q^{S,I}_{i,B}(t))+P^R_{i,B}(t)q^{R,S}_{i,B}(t), \nonumber \\
P^I_{i,B}(t+1)&=&P^I_{i,B}(t)(1-q^{I,R}_{i,B}(t))+P^S_{i,B}(t)q^{S,I}_{i,B}(t),  \\
P^R_{i,B}(t+1)&=&P^R_{i,B}(t)(1-q^{R,S}_{i,B}(t))+P^I_{i,B}(t)q^{I,R}_{i,B}(t). \nonumber
\end{eqnarray}
The first term on the right-hand side of the first equation of Eq. (\ref{eq:MC1}) is the probability that node $i$ is
remained as susceptible state. The second term stands for the probability that node $i$ is changed from refractory to
susceptible state. Similarly, we have the same explanation for the other equations of Eqs. (\ref{eq:MC1}) and (\ref{eq:MC2}).
Substituting Eqs. (\ref{eq:qsiA}) and (\ref{eq:qsiB}) into Eqs. (\ref{eq:MC1}) and (\ref{eq:MC2}), we obtain the microscopic
Markov dynamics as follows
\begin{eqnarray}\label{eq:theoryA}
P^S_{i,A}(t+1)&=&P^S_{i,A}(t)[(1-p_0)\prod_{l\in\Lambda_{i,A}}(1-\beta(t)P^I_{l,A}(t))\prod_{v\in\Lambda_{i,B}}(1-\beta_{ab}P^I_{v,B}(t))]+P^R_{i,A}(t)\delta,\nonumber\\
P^I_{i,A}(t+1)&=&P^I_{i,A}(t)(1-\gamma)+P^S_{i,A}(t)[1-(1-p_0)\prod_{l\in\Lambda_{i,A}}(1-\beta(t)P^I_{l,A}(t))\prod_{v\in\Lambda_{i,B}}(1-\beta_{ab}P^I_{v,B}(t))],\\
P^R_{i,A}(t+1)&=&P^R_{i,A}(t)(1-\delta)+P^I_{i,A}(t)\gamma. \nonumber
\end{eqnarray}
\begin{eqnarray}\label{eq:theoryB}
P^S_{i,B}(t+1)&=&P^S_{i,B}(t)[(1-p_0)\prod_{l\in\Lambda_{i,B}}(1-\beta(t)P^I_{l,B}(t))\prod_{v\in\Lambda_{i,A}}(1-\beta_{ab}P^I_{v,A}(t))]+P^R_{i,B}(t)\delta,\nonumber\\
P^I_{i,B}(t+1)&=&P^I_{i,B}(t)(1-\gamma)+P^S_{i,B}(t)[1-(1-p_0)\prod_{l\in\Lambda_{i,B}}(1-\beta(t)P^I_{l,B}(t))\prod_{v\in\Lambda_{i,A}}(1-\beta_{ab}P^I_{v,A}(t))],\\
P^R_{i,B}(t+1)&=&P^R_{i,B}(t)(1-\delta)+P^I_{i,B}(t)\gamma. \nonumber
\end{eqnarray}

Instead of getting the analytic solutions of Eqs. (\ref{eq:theoryA}) and (\ref{eq:theoryB}), we solve them by numerical
integration. We set the initial conditions as $P^S_{i,A}(0)=1.0$, $P^I_{i,A}(0)=0.0$, $P^R_{i,A}(0)=0.0$, $P^S_{i,B}(0)=1.0$,
$P^I_{i,B}(0)=0.0$, and $P^R_{i,B}(0)=0.0$. To conveniently compare the solutions with the numerical simulations in the section
{\it Results}, we use the same set of $\beta(t)$ for both the integration and numerical simulations. Fig. \ref{Fig:theory} shows
the results where the left and right panels are for the networks $\mathcal{A}$ and $\mathcal{B}$, respectively. In
Fig. \ref{Fig:theory}(b)-(d) and Fig. \ref{Fig:theory}(f)-(h), the solid curves represent the theoretical solutions while the
``circles" represent the numerical simulations. It is easy to see that the theoretical solutions are consistent with the
numerical simulations very well.

\subsection*{A theoretical analysis based on mean field theory}
To go deeper into the mechanism of the synchronized and mixed outbreak patterns, we try another theoretical analysis on mean field
equations. Let $s^A(t)$, $i^A(t)$ and $r^A(t)$ represent the densities of susceptible, infected, and refractory individuals at
time $t$ in network $\mathcal{A}$, respectively. Similarly, we have $s^B(t)$, $i^B(t)$ and $r^B(t)$ in network $\mathcal{B}$.
Then, they satisfy
\begin{eqnarray}\label{eq:sir}
 s^A(t)+i^A(t)+r^A(t)= 1, \quad
 s^B(t)+i^B(t)+r^B(t)=1.
\end{eqnarray}

According to the mean-field theory, we have the following ordinary differential equations
\begin{eqnarray}\label{eq:MF1}
\frac{ds^A(t)}{dt} &=& -p_0s^A(t)-\beta(t) \langle k_a\rangle s^A(t)i^A(t)-\beta_{ab}\langle k_{ab} \rangle s^A(t)i^B(t)+\delta r^A(t),  \\
\frac{di^A(t)}{dt} &=& p_0s^A(t)+\beta(t) \langle k_a\rangle s^A(t)i^A(t)+\beta_{ab}\langle k_{ab} \rangle s^A(t)i^B(t)-\gamma i^A(t),   \\
\frac{dr^A(t)}{dt} &=& \gamma i^A(t)-\delta r^A(t),
\end{eqnarray}
\begin{eqnarray}\label{eq:MF2}
\frac{ds^B(t)}{dt} &=& -p_0s^B(t)-\beta(t) \langle k_b\rangle s^B(t)i^B(t)-\beta_{ab}\langle k_{ab} \rangle s^B(t)i^A(t)+\delta r^B(t),  \\
\frac{di^B(t)}{dt} &=& p_0s^B(t)+\beta(t) \langle k_b\rangle s^B(t)i^B(t)+\beta_{ab}\langle k_{ab} \rangle s^B(t)i^A(t)-\gamma i^B(t),   \\
\frac{dr^B(t)}{dt} &=& \gamma i^B(t)-\delta r^B(t).
\end{eqnarray}
Specifically, we consider a case of single epidemic outbreak with extremely weak coupling, i.e. $p_0=0$ and $\beta_{ab}\approx 0$.
In the steady state, we have
\begin{eqnarray}\label{eq:steady}
\frac{ds^A(t)}{dt} = 0, \quad
\frac{di^A(t)}{dt} = 0, \quad
\frac{ds^B(t)}{dt} = 0, \quad
\frac{di^B(t)}{dt} = 0,
\end{eqnarray}
which gives
\begin{eqnarray}\label{eq:steady1}
s^A(t)=\frac{\gamma}{\beta(t) \langle k_a\rangle}, \quad
r^A(t) = \frac{\gamma}{\delta} i^A(t), \quad
s^B(t)=\frac{\gamma}{\beta(t) \langle k_b\rangle}, \quad
r^B(t) = \frac{\gamma}{\delta} i^B(t).
\end{eqnarray}
Substituting Eq. (\ref{eq:steady1}) into Eq. (\ref{eq:sir}) we obtain
\begin{eqnarray}\label{eq:steady2}
\frac{\gamma}{\beta(t) \langle k_a\rangle}+(\frac{\gamma}{\delta}+1)i^A(t)=1, \quad
\frac{\gamma}{\beta(t) \langle k_b\rangle}+(\frac{\gamma}{\delta}+1)i^B(t)=1.
\end{eqnarray}
The critical point can be given by letting $i^A(t)$ and $i^B(t)$ in Eq. (\ref{eq:steady2}) change from zero to nonzero, which
gives
\begin{eqnarray}\label{eq:critical}
\beta_c^A=\frac{\gamma}{\langle k_a\rangle}, \quad
\beta_c^B=\frac{\gamma}{\langle k_b\rangle}.
\end{eqnarray}
For the considered networks with $\langle k_a\rangle > \langle k_b\rangle$, we have $\beta_c^A < \beta_c^B$.

\subsection*{Cross-correlation measure}
In statistics, the Pearson correlation coefficient is a measure of the linear correlation between two variables.
If two time series $\{X_t\}$ and $\{Y_t\}$ have the mean values $\overline{X}$ and $\overline{Y}$, we can
define the correlation coefficient $r$ as follows
\begin{equation}\label{eq:Correlation coefficient}
r=\frac{\sum\limits_{t=1}^n(X_t-\overline{X})(Y_t-\overline{Y})}{\sqrt{\sum\limits_{t=1}^n
(X_t-\overline{X})^2\cdot\sum\limits_{t=1}^n(Y_t-\overline{Y})^2}}
\end{equation}
To analyze the correlations of the growth trends between the two time series, we investigates their cross-correlation $r(t)$
in a given window $w_t$ \cite{Podobnik:2008,Wang:2016}, i.e. $\{X_t,\ X_{t+1},\ \ldots, X_{t+w_t}\}$ and
$\{Y_t,\ Y_{t+1},\ \ldots, Y_{t+w_t}\}$. $\{X_t\}$ and $\{Y_t\}$ will share the same trend in the time interval $w_t$ when
$r(t)>0$ and the opposite growth trend when $r(t)<0$. In this work, we let the whole time series be one window, i.e. with
$w_t$ being the total evolutionary time $t$.

\section*{Acknowledgement}
We acknowledge the Centre for Health Protection, Department of Health, the Government
of the Hong Kong Special Administrative Region  and the USA National
Notifiable Diseases Surveillance System as digitized by Project Tycho for
providing data. We especially acknowledge Hern\'an A. Makse for supporting the computer resource
in the lab. This work was partially supported by the NNSF of China under Grant
Nos. 11135001, 11375066, 973 Program under Grant No.
2013CB834100, the Program for Excellent Talents in Guangxi Higher Education Institutions and
Guangxi Natural Science Foundation 2015jjGA10004.

\section*{Author contributions}
M.Z. and Z.L. conceived the research project. M.Z., M.Z., B.M. and Z.L.
performed research and analyzed the results.
M.Z., B.M. and Z.L. wrote the paper. All authors reviewed and approved
the manuscript.

\section*{Additional information}
Competing financial interests: The authors declare no competing
financial interests.
Correspondence and requests for materials should be addressed to
Z.L.~(zhliu@phy.ecnu.edu.cn).

\end{document}